\documentclass[twocolumn,prb,citeautoscript,superscriptaddress,a4paper,floatfix]{revtex4}
\usepackage[dvipdfm]{graphicx}
\newcommand{\feo}{Fe$_{2}$O$_{3}$}
\newcommand{\alumina}{Al$_{2}$O$_{3}$}
\newcommand{\silica}{SiO$_{x}$}
\newcommand{\etal}{\textit{et al.}}
\newcommand{\ie}{\textrm{i.e.}}
\newcommand{\eg}{\textrm{e.g.}}
\newcommand{\IV}{$I$-$V$}
\begin{document}
\title{Nonpolar resistance switching of \\ metal/binary-transition-metal 
oxides/metal sandwiches: \\ homogeneous/inhomogeneous transition of current distribution.}
\author{I. H. Inoue}
\affiliation{Correlated Electron Research Center (CERC),
    National Institute of Advanced Industrial Science and Technology (AIST),\\
    AIST Tsukuba Central 4, Tsukuba 305-8562, Japan.}
\homepage{http://staff.aist.go.jp/i.inoue/}
\author{S. Yasuda}
\affiliation{Nanotechnology Research Institute (NRI),\\
	AIST Tsukuba Central 2, Tsukuba 305-8568, Japan}
\affiliation{Department of Advanced Materials,
	University of Tokyo, Kashiwa 277-8581, Japan.}
\author{H. Akinaga}
\affiliation{Nanotechnology Research Institute (NRI),\\
	AIST Tsukuba Central 2, Tsukuba 305-8568, Japan}
\author{H. Takagi}
\affiliation{Correlated Electron Research Center (CERC),
    National Institute of Advanced Industrial Science and Technology (AIST),\\
    AIST Tsukuba Central 4, Tsukuba 305-8562, Japan.}
\affiliation{Department of Advanced Materials,
	University of Tokyo, Kashiwa 277-8581, Japan.}
\date{\today}
\begin{abstract}%
Exotic features of a metal/oxide/metal (MOM) sandwich, which will be the basis for a drastically innovative nonvolatile memory device, is brought to light from a physical point of view.
Here the insulator is one of the ubiquitous and classic binary-transition-metal oxides (TMO), such as \feo, NiO, and CoO.
The sandwich exhibits a resistance that reversibly switches between two states: one is a highly resistive off-state and the other is a conductive on-state.
Several distinct features were universally observed in these binary TMO sandwiches: namely, nonpolar switching, non-volatile threshold switching, and current--voltage duality.
From the systematic sample-size dependence of the resistance in on- and off-states, we conclude that the resistance switching is due to the homogeneous/inhomogeneous transition of the current distribution at the interface.
\end{abstract}
\maketitle
\section{INTRODUCTION}
For half a century, metal/oxide/metal (MOM) sandwich structures have been intensively examined.
Especially, it shows numerous interesting properties upon ``electroforming'' ---\,applying a voltage above a certain critical value to the sandwich to produce a permanent (nonvolatile) change in its electric properties \cite{kreynina,hickmott62,gibbons,chopra,simmons}.
These electroformed sandwich often exhibits a negative differential conductance (NDC) concomitant with electron emission, electroluminescence, and resistance switching \cite{dearnaley}.
Typical examples are \alumina-based \cite{hickmott00} and \silica-based \cite{ueno} sandwiches.
These phenomena were studied intensively until 1980s in a bid to put the sandwich to practical use.
Extensive reviews were given by Dearnaley \etal \cite{dearnaley}, Biederman \cite{biederman}, Oxley \cite{oxley}, and Pagnia \etal \cite{pagnia}.
In those reviews, it has been commonly stated that the most important facts are voids, dislocations, defects, and so on, which are, in a word, nonstoichiometry inevitable in every oxide thin film.

\begin{figure}[!htb]
	\centering
	\vspace{-0.7\intextsep}
	\includegraphics[width=0.38\textwidth,clip]{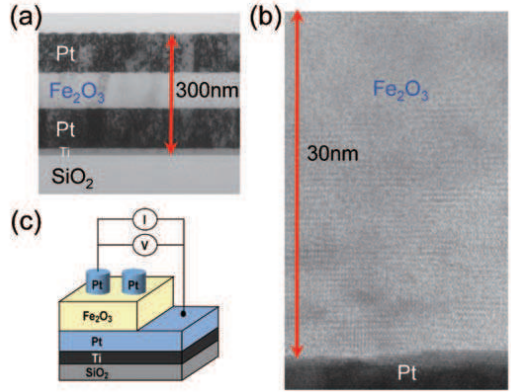}  
	\caption{
(a) Transmission electron microscope (TEM) cross-section image of a typical \feo-based sample used in the present measurements.
(b) Expanded view of the TEM image near the bottom interface of \feo\ and Pt, highlighting the \feo\ film
is in a polycrystalline state rather than in an amorphous state.
(c) Schematic view of the sample and the overall circuit arrangement used in the measurement of the electrical characteristics.
	}
	\label{device}
\end{figure}

Since around the year of 2000, there is a renewed interest in this area that was prompted by a new generation of
experimental \cite{beck,liu,stefanovich,hu,tulina,thurstans,baikalov,seo1,rohde,fors,szot,hamaguchi,dsjeong1,kmkim,yhyou,mjlee,hosoi,ysato,sthsu} and theoretical works \cite{gravano,prl,oka,apl,shjeon,dsjeong2}, which have rekindled the long-running controversy on the mechanism behind the resistance switching phenomena of the electroformed sandwich.
We demonstrate in the present work that binary transition-metal oxide (TMO)-based sandwiches show unique characteristics such as nonpolar switching, non-volatile threshold switching, and current--voltage duality, which can be assigned to a different category from those of the long-time-studied \alumina-based and \silica-based sandwiches.
The samples we have fabricated are Pt/\feo/Pt, Pt/NiO/Pt, and W/CoO/Pt sandwiches as shown in Fig.\,\ref{device}.
\section{EXPERIMENTAL}
Three kinds of heterostructures of Pt/\feo/Pt/Ti, Pt/NiO/Pt/Ti, and CoO/Pt/Ti are grown by conventional radio-frequency magnetron sputtering on commercially-available thermally-oxidised single-crys\-tal\-line Si substrates.
Stoichiometric \feo, NiO and CoO commercial targets are used.
The film growth of Ti (20\,nm thick) and Pt (100\,nm thick) is achieved in an argon discharge at a pressure of 0.5\,Pa and 0.3\,Pa, respectively, at room temperature.
The film growth of \feo, NiO, and CoO is done in an Ar 96\,\% and O$_{2}$ 4\,\% discharge at a total pressure of 0.67\,Pa with a substrate temperature of 300$^{\circ}$C\@.
After depositing each of these oxides, the film is cooled to room temperature maintaining the same Ar/O$_{2}$ flow and total pressure.

It should be noted here that each of the expressions ``\feo'', ``NiO'', and ``CoO'' in this paper denotes its nominal composition, because we have not examined the stoichiometry of the oxide films.
The chemical composition is possibly different from the nominal value.

The radio-frequency power is 200\,W except for the  Pt deposition (100\,W) with corresponding growth rates of $\sim$10\,nm/min (Ti and Pt), $\sim$3.2\,nm/min (\feo), $\sim$6\,nm/min (NiO), and $\sim$7\,nm/min (CoO)\@.
The top Pt electrodes of the \feo- and NiO-based samples are patterned by the standard photolithography with Ar ion milling to the shape of 200, 100, and 60\,$\mu$m diameter column.
It should be also noted that the CoO film grown in this work is much more conductive than the films of \feo\ and NiO.
Therefore, instead of depositing Pt as a top electrode, a mechanical contact to a tungsten needle as described below is used for the CoO film in order to reduce the total current.

The samples are characterised using a transmission electron microscopy (TEM) of H-9000NAR (Hitachi) operated at 300\,kV\@.
Samples for the TEM observation are prepared by mechanical grinding, dimpling to a thickness of less than 50\,nm, and then ion-beam thinning to electron transparency.

A conventional microprobe station with tungsten needle contacts ($10\,\mu$m diameter) is used for the electrical contact.
The needle is mechanically placed either on the Pt electrodes of the \feo- and NiO-based samples, or on the CoO film directly with the utmost care that the tip does not scratch or perforate the oxide surface.
Bias polarity is defined with reference to the bottom Pt electrode.
The Agilent 4156C Precision Semiconductor Parameter Analyser is used to measure the d.c.\ electrical properties of the sample.
A current $I$ or voltage $V$ limitation (compliance) is set for the $V$- or $I$-sweep measurement, respectively, to prevent breakdown and destruction of the samples.
All the measurements are performed in air in the dark at room temperature.

\begin{figure}[!htb]
	\centering
	\includegraphics[width=0.48\textwidth,clip]{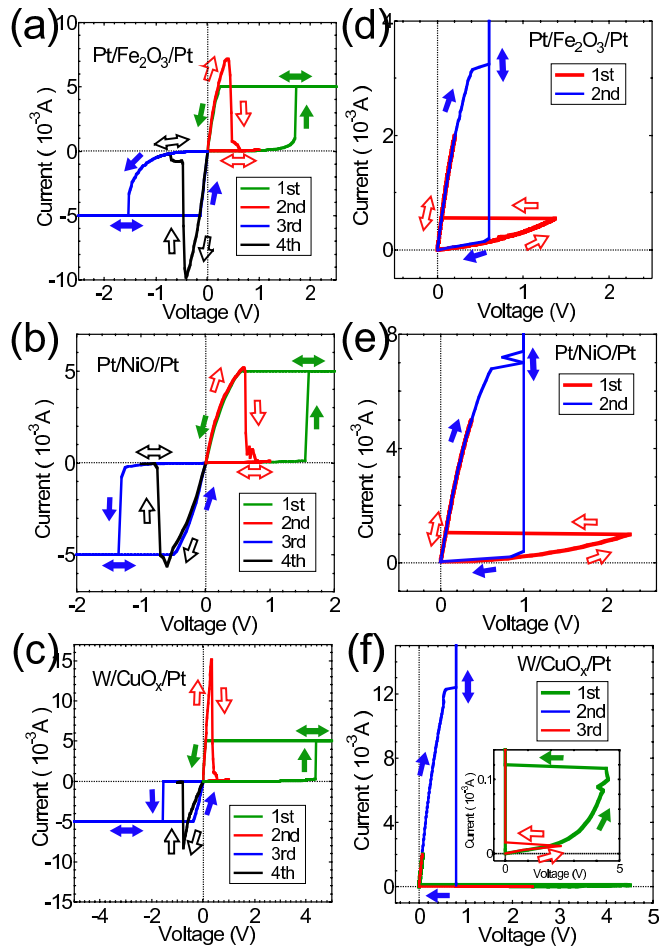}
	\caption{%
\IV\ characteristics of Pt/\feo/Pt (a),  Pt/NiO/Pt (b), and W/CoO/Pt (c) for $V$-sweep measurements. 
Thickness of the oxide is 100\,nm (\feo), 60\,nm (NiO), and 70\,nm (CuO).
Diameter of the top Pt electrode is 60\,$\mu$m for the \feo\ and NiO samples, while the top W electrode ($10\,\mu$m diameter) is mechanically placed for the CuO sample.
The arrows indicate the direction of the $V$ sweep. 
Only for the first and third measurements, we set $I$-compliance of $\pm\,5$\,mA. 
\IV\ characteristics of Pt/\feo/Pt (d),  Pt/NiO/Pt (e), and W/CoO/Pt (f) for $I$-sweep measurements. 
The arrows indicate the direction of the $I$ sweep. 
$V$-compliance is $0.6$\,V for Pt/\feo/Pt and W/CoO/Pt, and is 1\,V for Pt/NiO/Pt. 
The inset of (f) shows the blow-up. 
Result of the third sweep is additionally plotted only for this sample, because it was notably different from the first one.
	}
	\label{linearIV}
\end{figure}

\section{RESULTS AND DISCUSSIONS}
\subsection{Threshold switching and \IV\ duality}
Remarkable features are found in the current--voltage \IV\ characteristics of the three samples plotted in Fig.\,\ref{linearIV}.
We have performed d.c.-voltage sweep measurements (Figs.\,\ref{linearIV}(a), \ref{linearIV}(b) and \ref{linearIV}(c)), and d.c.-current sweep measurements (Figs.\,\ref{linearIV}(d), \ref{linearIV}(e) and \ref{linearIV}(f)).
All the samples show an abrupt change of resistance from the highly resistive off-state to the more conductive on-state.
This is called ``set'' which is seen in the first and third scans.
On the other hand, ``reset'' is seen in the second and fourth scans, as another sharp resistance change from the on-state to the off-state.

\begin{figure}[!htb]
	\centering
	\includegraphics[width=0.48\textwidth,clip]{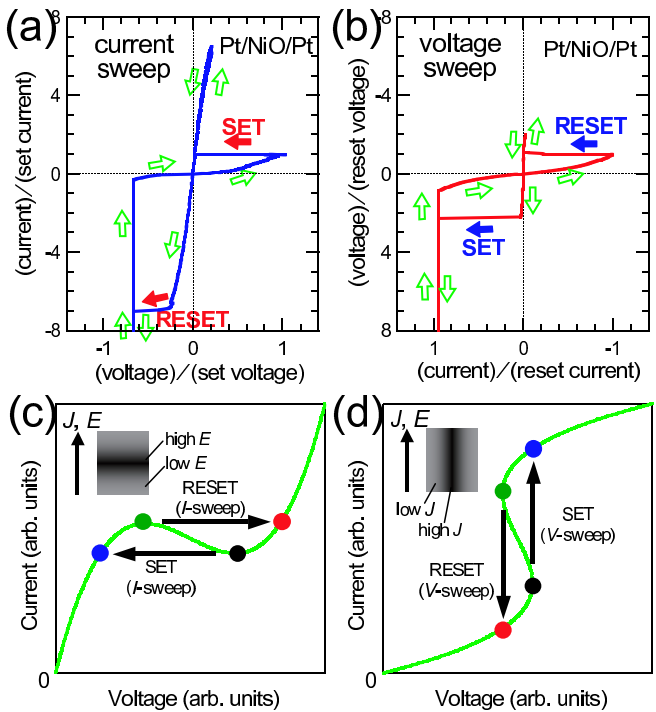}
	\caption{%
\IV\ characteristics of Pt/NiO/Pt for $I$-sweep measurement (a) and $V$-sweep measurement (b).
The axes are normalised to the set threshold value for (a), and to the reset threshold value for (b).
(c) N-type negative differential conductance generally due to the self-organised spatial pattern formation of electric field domains as illustrated in the inset, where $J$ is the current density and $E$ is the electric field.
The current is a single-valued function of the voltage, but the voltage is a multi-valued one; \ie, the `set' and `reset' are seen only in the $I$-sweep measurement.
(d) S-type negative differential conductance generally due to the current filamentation as shown in the inset.
The `set' and `reset' are seen only in the $V$-sweep measurement.
	}
	\label{IVscale}
\end{figure}

An intriguing property that we notice in the \IV\ characteristics is the duality relation of $I$ and $V$.
If the off-state can be fitted by a certain function $f$, \ie, $I$\,$=$\,$f(V)$, then the on-state can be fitted by $V$\,$=$\,$A f^{-1}(BI)$, where $A$ and $B$ are scale constants. 
The sharp `set' and `reset' of the resistance can be observed both in the $I$- and $V$-sweep measurements in a remarkably similar fashion.
In fact, we can hardly tell which is $I$ or $V$ axis, if only the raw data are plotted with neither scale nor label on the axes (see Figs.\,\ref{IVscale}(a) and \ref{IVscale}(b) where $I$ vs.\ $V$ and $V$ vs.\ $I$ are shown).

The discontinuous resistance switching might be ascribed to the non-equilibrium phase transition that is induced when sufficiently high electric field is applied or large current is injected \cite{ridley,jager,scholl}.
This type of electric instability is generally followed by the N-type NDC due to an inhomogeneity of the electric field (Fig.\,\ref{IVscale}(c)) or followed by the S-type NDC due to a current filamentation (Fig.\,\ref{IVscale}(d)).
In those cases, the so-called ``threshold switching''\cite{adler} is observed in either of the $V$- or $I$-sweep measurement.
However, both the N-type and the S-type NDCs due to such electric instabilities \cite{scholl} are volatile as apparent from Figs.\,\ref{IVscale}(c) and \ref{IVscale}(d). 
Thus, in this sense, the `set' and `reset' switchings with the \IV\ duality and large hysteresis observed in our samples are fairly unique, and distinctly different from the conventional threshold switching.
Nevertheless, we would like to use the term ``threshold switching'' for the observed sharp `set' and `reset' switchings of our samples, simply as a counterpart of another well-known non-volatile resistance switching, which is ``continuous'' or ``branch'' switching as seen, for instance, by altering a Schottky barrier height/width arising from the reversible electron movement \cite{pan,sawa,smits}.

\subsection{Nonpolar switching of resistance}
Another remarkable feature of our results is the intriguing nonpolar nature of the resistance switching.
Once we `set' the sample by applying a positive $V$ or $I$, then it is `reset' by applying another positive $V$ or $I$.
This is a typical behavior of the unipolar switching. 
In general, the unipolar switching can be seen for either positive or negative polarity of $V$ or $I$, our samples show `set' and `reset' when alternating the polarity of $V$ or $I$ as seen in bipolar switching.
Thus, the unipolar and bipolar actions are coexisting in a quite unique fashion, and we propose that this switching should be better termed by ``nonpolar'' action rather than unipolar or bipolar.

The mechanism of this nonpolar behaviour requires more detailed studies to be clarified, and quite interesting and suggestive study with regard to the polarity has been reported recently by Hosoi \etal\cite{hosoi}
They have succeeded to change a bipolar switching of a TiN/TiON/TiN sandwich into a unipolar one (only in positive $V$ side) by simply connecting a load resistance in series.
So far, it was suggested that bipolar switchings of MOM structures could be due to an explicit asymmetry \cite{baikalov} such as a poling (training) of the oxide produced by large electric pulses.
However, Hosoi \etal\ has demonstrated that the bipolar switching seen in an apparently symmetric sandwich can be turned into a unipolar one by introducing the external asymmetry.
This might mean that the bipolar and unipolar actions are originally identical, and geometric and/or electronic asymmetries of the sandwich device can change the essential ``nonpolar'' action into the apparent bipolar and unipolar actions.
When we consider the mechanism of the resistance switching, this essential nonpolarity and its relation to the external asymmetry is important, since it may be unnecessary to propose different mechanism for each type of resistance switchings.

\begin{figure}[!htb]
	\newpage  
	\centering
	\includegraphics[width=0.49\textwidth,clip]{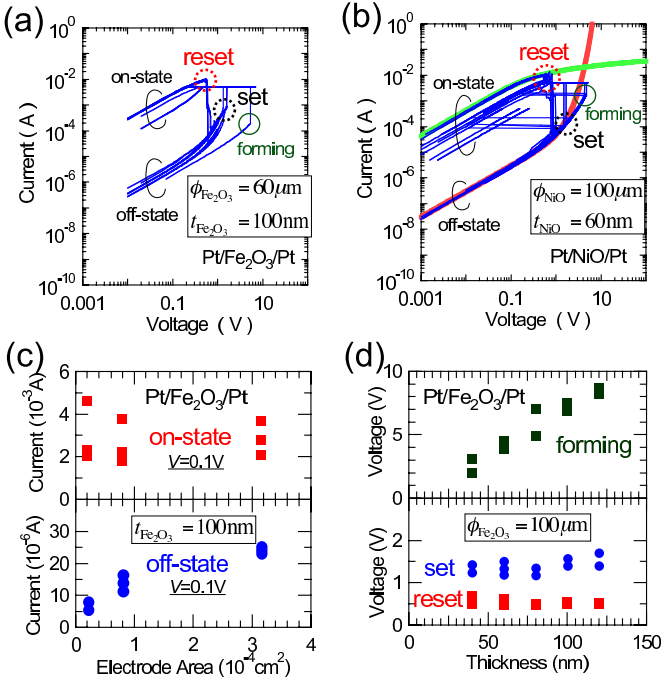}
	\caption{%
(a) \IV\ characteristics obtained by 27 successive $V$-sweep measurements of a single Pt/\feo/Pt sample. 
`Reset' and `set' voltages are $\sim$0.5\,V, and $\sim$1.4\,V, respectively. 
`Set' voltage of the first sweep (forming) is larger ($\sim$5.3\,V) than the followings.
(b) Same as (a) but obtained by 58 successive $I$-sweep and $V$-sweep measurements of four Pt/NiO/Pt samples.
The `reset', `set', and `forming' voltages are $\sim$0.7\,V, $\sim$1.6\,V, and $\sim$4.4\,V, respectively.
Thick red line corresponds to $I/I_\mathrm{off}$=$\sinh(V/V_\mathrm{off})$, where $I_\mathrm{off}$=$16$\,$\mu$A, and $V_\mathrm{off}$=$0.56$\,V.
Thick green line corresponds to $V/V_\mathrm{on}$=$\sinh(I/I_\mathrm{on})$, where $I_\mathrm{on}$=$4.8$\,mA, and $V_\mathrm{on}$=$0.11$\,V.
(c) $I$ of the on-state (top) and off-state (bottom) at $V$=$0.1$\,V for the $V$-sweep measurements of Pt/\feo/Pt samples, plotted against the area of top Pt electrode.
(d) `Reset' and `set' voltages (bottom) as well as forming voltage (top) for $V$-sweep measurements of Pt/\feo/Pt samples, plotted against the thickness of \feo.
     }
	\label{logIV}
\end{figure}

\subsection{Homogeneous/inhomogeneous transition of current distribution ---\,a ``faucet'' model}
Figure\,\ref{logIV}(a) shows a collection of \IV\ curves obtained by 27 successive $V$-sweep measurements for a single Pt/\feo/Pt sample, and Fig.\,\ref{logIV}(b) shows all the results of 58 measurements of $I$- and $V$-sweep for four Pt/NiO/Pt samples with the same geometry.
For all these measurements, the $V$-sweep rate and the $I$-sweep rate is kept constant ($\sim\!80$\,mV/sec and $\sim\!80\,\mu$A/sec).
We immediately notice that $I$ is always linear in $V$ at low fields.
By comparing the \IV\ curves of the different sweeps, it is clear that the off-state resistance is quite reproducible with a slight scattering of around $30$\,k$\Omega$.
In contrast, the on-state resistance shows a considerable variation ranging from 30 to 200\,$\Omega$ for Pt/\feo/Pt, and from 20 to 500\,$\Omega$ for Pt/NiO/Pt.
This scattering of on-state resistance values suggests the presence of inhomogeneous distribution of the current paths that forms in the on-state.

In order to construct a clearer image of the resistance switching, the systematic size-dependence was investigated for Pt/\feo/Pt.
In Fig.\,\ref{logIV}(c), the current at a constant voltage of 0.1\,V is plotted as a function of the top electrode area.
The thickness of samples used in Fig.\,\ref{logIV}(c) is kept at 100\,nm.
(That is, the data represents an area-dependence of the inverse resistance.)
In the off-state shown in the bottom panel, the current is almost proportional to the area of the top Pt electrode.
This indicates that, in the off-state, the current flows homogeneously over the electrode area.
Meanwhile, the on-state current does not depend on the area of the top Pt electrode as shown in Fig.\,\ref{logIV}(c)\,(top).
We therefore conclude that the `set'/`reset' switchings correspond to a homogeneous (off-state) to inhomogeneous (on-state) change of the conduction.

The relationship between the `set'/`reset' voltages and the thickness of the oxide indicates another important feature.
As apparent from Fig.\,\ref{logIV}(d)\,(bottom), both the `set' and `reset' voltages do not depend on the thickness of the oxide, and are almost constant.
This means the total bias voltage is mainly applied at a high resistance region, whose thickness does not depend on the thickness of the whole oxide.
It is natural that the region can be the metal/oxide interface as discussed later.
In marked contrast to the `set'/`reset' voltages, the voltage required for the initial electroforming increases linearly with the thickness of \feo, as shown in the top panel of Fig.\,\ref{logIV}(d), indicating that the electric field inside the bulk \feo\ is the controlling factor of the forming.
Once the electroforming is completed, it is unlikely that the conductive bulk region \cite{simmons} contributes to the switching effectively.
This also underpins that the `set'/`reset' voltages are dominantly applied to the interface rather than the bulk region.

\begin{figure}[!htb]
	\newpage
	\centering
	\includegraphics[width=0.49\textwidth,clip]{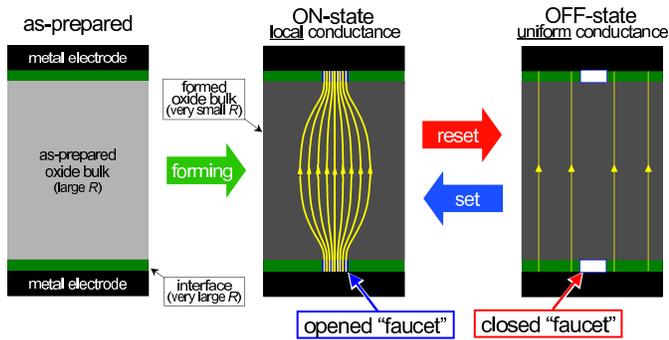}
	\caption{%
Schematic pictures of the forming, set and reset: an electric faucet model.
By applying a large electric field to the as-prepared sample (left), the whole bulk region of the oxide become much more conducting, as well the ``electric faucet'' opens in one or both interface high-resistance region(s) to connect the metal electrode(s) to the oxide bulk (middle).
This procedure is called the ``electroforming'' or simply ``forming''.
The thin yellow lines in the middle and right panels illustrate the schematic electric current distribution.
When a faucet (not necessary to be the both ones) is opened on the on-state (as colored blue), the current flows through the opened faucet rather than through the high-resistance interface.
Thus, the electric conduction becomes inhomogeneous in the on-state (middle).
Meanwhile, when the opened faucet is closed (as colored white), the current prefers to go through the whole interface, so that the electric conduction becomes  homogeneous (right).%
	}
	\label{schematics}
\end{figure}

A natural scenario to explain these results is a formation of a narrow conducting path or ``filament'' from the bottom to the top electrode \cite{gibbons,dearnaley,pagnia,rohde,szot,dsjeong1}, and the disconnection/reconnection of the filament at `reset'/`set' which occurs at the interfaces \cite{filament_rupture}.
However, considering that the off-state current seems to flow homogeneously over the electrode area within our experimental accuracy,
it is rather feasible to assume the bulk of oxide in the off-state would be an almost uniformly conducting matrix \cite{filament_remnant}.
Therefore, we propose a model based on an ``electric faucet'' formed at the interface barrier as schematically shown in Fig.\,\ref{schematics}.
The model postulates that the bulk of oxide would become a conductive region \cite{yhyou}, after the electroforming, with \eg, a lot of conductive filaments, domains, or anything conductive, but the total current flow could be controlled by the faucet located at the high resistance interface.
The faucet can be regarded a tip of a particular filament reached to the electrode, or a conductive grain/domain penetrating the interface barrier.
The area of the faucet must be much smaller than the whole area of the electrodes, and the resistance in the on-state, where current flows through the open faucet, is almost area-independent.
The observed variation of the on-state current may thus occur due to the different sizes of the formed faucet through successive switching cycles.
In the off-state, the current flows almost homogeneously over the highly resistive interface.
It should be noted that the on/off of faucets does not need to occur simultaneously at both interfaces. 
If at least one faucet exits in either of the interfaces, the above scenario will hold.

\subsection{Possible mechanisms for opening/closing of the faucets}
To consider the physics behind the opening and closing of the faucet, we have focused our attention on the `reset' voltages in Fig.\,\ref{logIV} that are remarkably reproducible (for fixed $V$-sweep rate) in spite of the large variation of the on-state current.
One possible explanation is to assume that the on-state current is scaled by the area of faucet; \ie, there is a critical current density $J_{\rm c}$ to drive the `reset' switching. 
Although there has been no definitive experimental evidence on the total area of faucet(s), if we assume it  300\,nm in diameter, the critical current density for `reset', $J_{\rm c}$, of a typical sample of our Pt/\feo/Pt sandwich can be estimated to be $10^{7}$\,A/cm$^{2}$ \cite{jc_estimation}.

This value is quite noteworthy, because a macroscopic material transport known as electromigration becomes significant \cite{lloyd,pierce} in such a high current density environment.
In general, electromigration forms nanocracks preferably where the materials transport is inhomogeneous and where the activation energy for diffusion is reduced compared to that of the bulk \cite{tu}. 
In our sandwich samples, such a place is most probably an interface between the oxide and the metal electrode \cite{pippel}.
Once a nanocrack is formed there, the current density through or around the nanocrack starts increasing.
When the current density is increased, the nanocracks grows further, and thus the current gets confined in this region at a rapidly accelerating rate \cite{schmidt,martel}.
We think this high-current-density region conforms to the faucet.
In perovskite-type oxides, electromigration of oxygen atoms is known to occur at much smaller current densities of $10^{3}$--$10^{6}$\,A/cm$^{2}$ \cite{huerth,ghosh,moeckly}, since oxygen is particularly mobile and the activation energy for diffusion is typically less than 1\,eV.
Indeed, electromigration in transition-metal oxides has been of concern, because the oxygen movement seriously limit the future electronic application of these oxides.

However, less has been understood about the electronmigration in complex oxides.
Even in the conventional metals, though it has been studied for a long time, there still remain fundamental controversies on the physical origins \cite{lloyd,verbruggen}.
Thus, here we focus only on the dynamics of the oxygen electromigration studied on LaNiO$_{3-\delta}$ \cite{ghosh} and on YBa$_{2}$Cu$_{3}$O$_{7-\delta}$ \cite{moeckly}.
The reports have revealed that the resistance change, $\Delta R$, due to the oxygen electromigration follows a streched-exponential dependence of time $t$, \ie, $\Delta R\,\propto\,1-e^{-(t/\tau)^{\beta}}$, for both the `set' and `reset' processes.
$\tau$ and $\beta$ are physical parameters.
According to the reports, the resistance change below a specific temperature ($\sim$350\,K) is quite slow (typically $\sim10^{4}$\,sec).
However, as the temperature increases, $\tau$ decreases exponentially for both `set' and `reset' switching \cite{ghosh}, and the time scale shows a drastic drop in the magnitude.
It is reasonable to imagine that the flux of electrons in the small faucet region may induce considerable local heating, which may raise the local temperature up to $\sim$1000\,K \cite{schmidt,martel}, apart from the conventional Joule heating \cite{gibbons,rohde,ysato,dsjeong2}.
This leads to the instantaneous resistance change \cite{hosoi} at the small faucet.

The above scenario is based on a semi-classical oxygen electromigration \cite{baikalov,szot,shjeon} with large-scale material displacement.
That is, as seen in the conventional metals, the `set' process heals defects, while the `reset' process increases the defects and disorder.
However, we have to assume high local temperature, and in such a high temperature atmosphere the equilibrium of oxygen activity is quite altered, resulting in the change of chemical composition by itself.
The changes could be significant especially at the interface due to the Gibbs' adsorption \cite{ricoult}.
Moreover, in several binary metal oxides with metal electrode, the averaged free enthalpy of oxygen segregation is close to the free reaction enthalpy of oxide precipitation from the metal \cite{pippel}, and it has been actually demonstrated that the current density of $10^{7}$\,A/cm$^{2}$ induces intriguing local oxidation \cite{schmidt,martel}. 
Accordingly, in addition to the possible ``direct effect'' of the oxygen electromigration \cite{baikalov,szot,shjeon}, another and preferable mechanism of the resistance switching in our samples is the ``local chemical reaction'', \ie, oxidation and reduction, at the non-equilibrium interface between the metal electrode and the oxide matrix.

In the last place, we would like to comment briefly on a challenging issue: the effect of electron correlations to the mechanism.
Some recent works have suggested that a local electron interaction can enhance the electromigration \cite{huerth,molen}, and our samples, transition-metal oxides, are typical examples of the strongly correlated materials.
In conventional metals such as gold, electromigration causes the global change of the structure, however in the transition-metal oxide, it seems to be only concomitant with a local structural change within a single domain \cite{huerth,domain_tunneling_model}.
Then, what occurs in the small domain could be regarded as a naive stoichiometry control due to the local oxidation/reduction, which should be influenced by the change of the local electronic states and the electron correlations may play some roles \cite{quintero}. 
Although more detailed and systematic studies are necessary to unravel the true nature, the concept is quite stimulating since it would imply a new class of ``correlated electron devices'', which have been much sought after in the long history of the research field.

\section{SUMMARY}
We have realised the nonvolatile resistance switching in strongly correlated binary TMO-based sandwiches, which is essentially different from that of the metal/conventional-oxide/metal sandwiches.
The $I$- and $V$-controlled abrupt `set'/`reset' switchings are all unipolar and bipolar simultaneously; we thus call this switching ``nonpolar''.
Moreover, the role of $I$ seem to be interchangeable with that of $V$, as indicated by an observed novel and intriguing \IV\ duality.
Those features may allow for ingenious functionalities when these are used as future nonvolatile memory devices.
The current flows uniformly in the off-state, while in the on-state the current is localised.
In addition, the `set'/`reset' voltages do not depend on the thickness of the oxide.
Thus, the `set'/`reset' switching is associated with the homogeneous/inhomogeneous transition of the current distribution; \ie, the ``electric--faucet'' turns on and off at the high-resistance interface to regulate the current flow.
These results provide a key to elucidate the mechanism of the resistance switching.
As well, they cast an important question to be tackled from a technological point of view: a complete control of the faucet size and miniaturization.
This is a challenging problem for realizing a new electronic device using this resistance switching phenomenon.

\section*{ACKNOWLEDGMENTS}

A part of this work was conducted in AIST Nano-Processing Facility.
We would like to thank H. Akoh, T. Fujii, K. Fujiwara, Y. Hosoi, M. Kawasaki, T. Manago, Y. Ogimoto, M. J. Rozenberg, M. J. S\'{a}nchez, A. Sawa, H. Shima, Y. Tamai, Y. Tokura, and M. Yamazaki for their support and useful discussions.
Financial support was partially provided by New Energy and Industrial Technology Development Organization (NEDO), and Grant-in-Aid for Scientific Research from MEXT, Japan.


\begin{thebibliography}{50}
\bibitem{filament_rupture}
If we assume the filament is ruptured by Joule heating, it should occur in the middle of the filament \protect\cite{dsjeong2}, since the metal electrodes work as a large heat sink.
\bibitem{filament_remnant}
Although it is not obvious within our experimental accuracy, if we assume that only a few ``disconnected filament(s)'' would remain in the bulk of oxide in the off-state as narrow but dominant local current paths, the off-state current could not show such an electrode-area dependence as shown in Fig.\,\ref{logIV}(c).
\bibitem{jc_estimation}
Since we have assumed the resistance of the oxide bulk is small enough to be negligible, the resistance change ratio of $\sim$10$^{3}$ indicates almost all the on-state current flows through the faucet.
\bibitem{domain_tunneling_model}
The situation has been assumed implicitly in Ref.\,\protect\onlinecite{prl} and Ref.\,\protect\onlinecite{apl}, suggesting the `small domains' near the interface behave as faucets which turn on and off by the electron correlations.
\end{thebibliography}
\end{document}